\documentclass[10pt,a4paper]{book}
\usepackage{udineHyderabad}
\usepackage{epsfig}
\begin{document}

\talktitle{Numerical modelling \\of quantum statistics in high-energy
physics}

\talkauthors{Oleg V. Utyuzh\structure{a},
             Grzegorz Wilk\structure{b},
             Zbigniew W\l odarczyk\structure{b}}

\authorstucture[a]{The Andrzej So\l tan Institute for Nuclear Studies,
                   Nuclear Theory Department,\\
                   ul. Ho\.za 69; 00-681 Warszawa, Poland; emails:
                   utyuzh@fuw.edu.pl, wilk@fuw.edu.pl}

\authorstucture[b]{Institute of Physics,
                   \'Swi\c{e}tokrzyska Academy,\\
                   ul. \'Swi\c{e}tokrzyska 15, 25-406 Kielce, Poland,
                   e-mail: wlod@pu.kielce.pl}

\shorttitle{Numerical modelling of quantum statistics}

\firstauthor{O.V.Utyuzh}

\begin{abstract}
Numerical modelling of quantum effects caused by bosonic or fermionic
character of secondaries produced in high energy collisions of
different sorts is at the moment still far from being established. In
what follows we propose novel numerical method of {\it modelling}
Bose-Einstein correlations (BEC) observed among identical (bosonic)
particles produced in such reactions. We argue that the most natural
approach is to work directly in the momentum space of produced
secondaries in which the Bose statistics reveals itself in their
tendency to bunch in a specific way in the available phase space.
Fermionic particles can also be treated in similar fashion.
\end{abstract}

\begin{figure}[ht]
\hspace{1.45cm}
  \begin{minipage}[ht]{104mm}
    \centerline{
        \epsfig{file=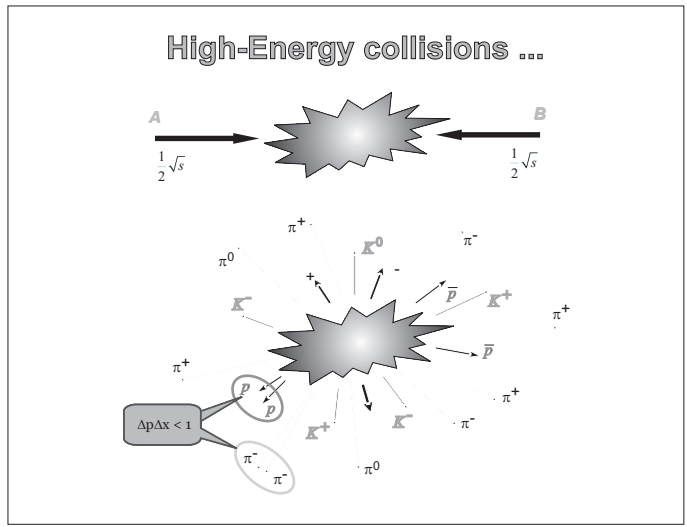, width=90mm}
               }
  \end{minipage}
  \caption{
\footnotesize {Schematic view of high energy collision resulting in
production of many particles of different statistics.}}
  \label{fig:Fig1}
\end{figure}
\begin{figure}[ht]
\hspace{1.45cm}
  \begin{minipage}[ht]{104mm}
    \centerline{
        \epsfig{file=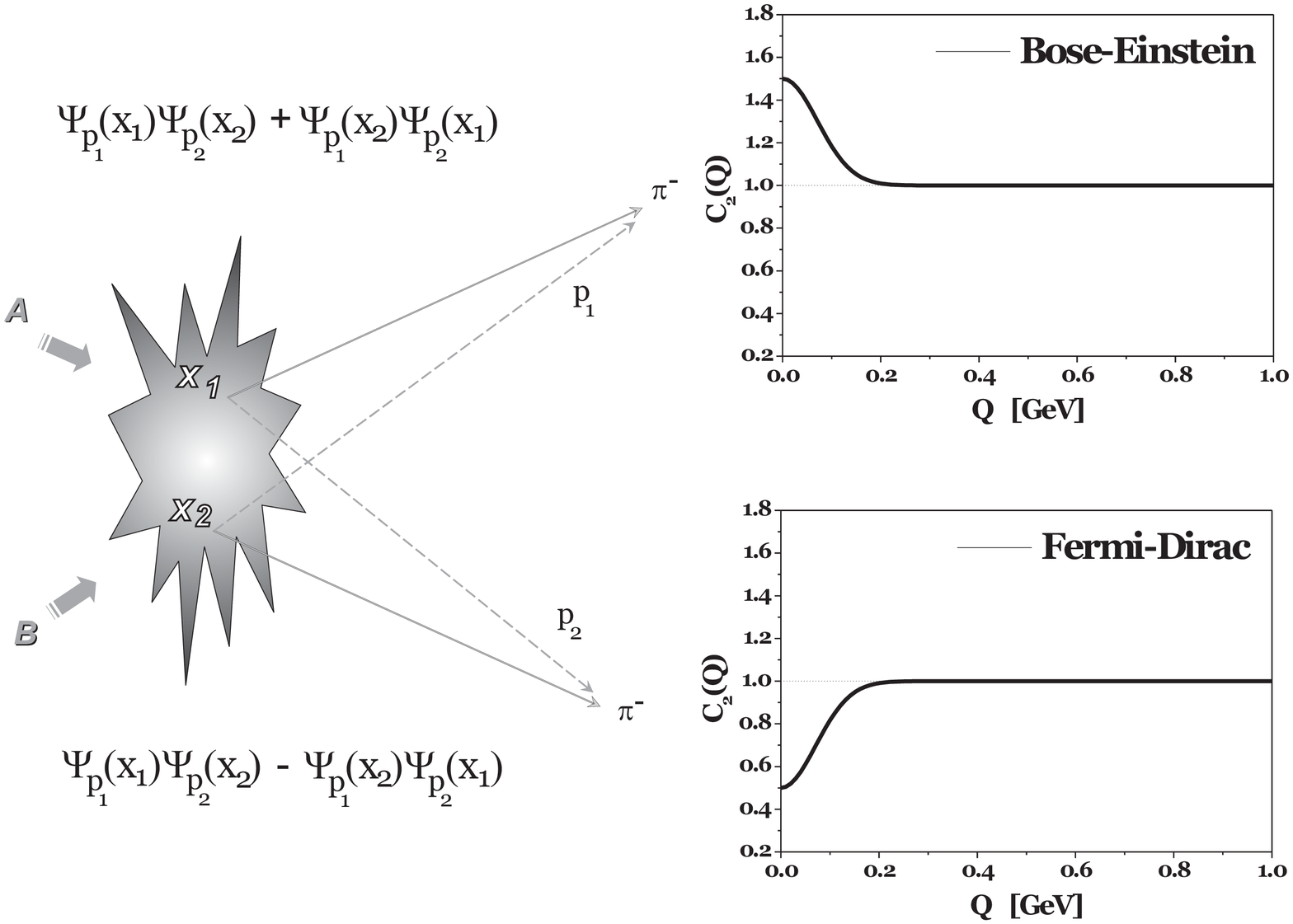, width=90mm}
               }
  \end{minipage}
  \caption{
\footnotesize {Example of BE and FD statistics (left panel) and two
particle correlation functions they lead to (right panels).}}
  \label{fig:Fig2}
\end{figure}

The multiparticle production processes consist substantial part of
the high energy collisions and are of considerable theoretical
interest. Unfortunately their description is so far available only by
means of numerical Monte Carlo codes based to some extent on modern
theoretical ideas but otherwise remaining purely phenomenological
\cite{GEN}. They are build in such manner as to describe as close as
possible the complicated final state of such reaction, cf., Fig.
\ref{fig:Fig1}. However, using classical (positive difined)
probabilities as their basic tool such MC codes cannot directly
describe some observed features, which are connected with
Bose-Einstein (BE) (or Fermi-Dirac (FD)) statistics of produced
secondaries (in case they are identical and occupy almost the same
parts of the phase space defined by the uncertainty relation, see
Fig. \ref{fig:Fig1}). When two (or more) identical particles of the
same kind are observed, their common wave function should be
symmetrized (for BE statistics) or antisymmetrized (for FD
statistics) what results in characteristic shapes of (two particle,
for example) correlation function $C_2(Q=|p_1-p_2|) =
N(p_1,p_2)/[N(p_1)\cdot N(p_2)]$, cf., Fig. \ref{fig:Fig2}.


Referring for details of Bose-Einstein correlations (BEC) to the
literature (cf., for example, \cite{BEC} and references therein) let
us concentrate here directly on the problem of their {\it proper
numerical modelling}, i.e., such in which the bosonic character of
secondaries produced in hadronization process are going to be
accounted for from the very beginning. This problem was so far
considered only in \cite{OMT} (using statistical approach based on
information theory approach, cf., however, also \cite{ZAJC}). All
other approaches, which claim to model BEC numerically \cite{modBEC},
simply add to the outcomes of existing MC codes \cite{GEN} some {\it
afterburners}, which modify them in a suitable way to be able fit the
BEC data. Such approach inevitably leads to such unwanted features as
violation of energy-momentum conservation or changes in the original
(i.e., obtained directly form MC code) multiparticle spectra.

In \cite{UWW} we have proposed afterburner free from such unwanted
effects. It was based on different concept of introducing quantum
mechanical (QM) effects in the otherwise purely probabilistic
distributions then those proposed in \cite{QUANT}. Namely, each MC
code provides us usually with a given number of particles, each one
endowed with either $(+)$ or $(-)$ or $(0)$ charge and with well
defined spatio-temporal position and energy-momentum. But experiment
provides us information on only the first and last characteristics.
The spatio-temporal information is not available directly. In fact,
the universal hope expressed in \cite{BEC,modBEC} is that precisely
this information can be deduced from the previous two via the
measured BEC. Our reasoning was as follows: $(i)$ BEC phenomenon is
of QM origin therefore one has to introduce in the otherwise purely
classical distributions provided by MCG the new element mimicking QM
uncertainties; $(ii)$ it cannot be done with energy-momenta because
they are measured and therefore fixed; $(iii)$ the next candidate,
i.e., spatio-temporal characteristics can be changed but it was
already done in \cite{QUANT,modBEC}; $(iv)$ one is thus left with
charges and in \cite{UWW} we have simply assigned (on event-by-event
basis) new charges to the particles from MCG conserving, however, the
original multiplicities of $(+/-/0)$. This has been done in such way
as to make particles of the same charge to be located maximally near
to each other in the phase space exploring for this natural
fluctuations in spatio-temporal and energy-momentum characteristic of
the outcome of MCG. The advantages of such approach are: $(a)$
energy-momentum is automatically conserved and multiparticle
distributions are not modified and $(b)$ it is applicable already on
the level of each event provided by MCG (not only, as some of
propositions of \cite{modBEC} only to all events). However, the new
assignment of charges introduces a profound change in the structure
of the original MCG. Generally speaking (cf. \cite{UWW} for details)
it requires introduction of bunchings of particles of the same
charge.

This observation will be the cornerstone of our new proposition. Let
us first remind that idea of bunching of particles as quantum
statistical (QS) effect is not the new one \cite{QS}. It was used in
connection with BEC for the first time in \cite{GN} and then was a
basis of the so called {\it clan model} of multiparticle
distributions leading in natural way to their negative binomial (NB)
form observed in experiment \cite{NBD}. It was then again introduced
in the realm of BEC in \cite{BSWW} and \cite{OMT,ZAJC}. Because our
motivation comes basically from \cite{OMT} let us outline shortly its
basic points. It deals with the problem of how to distribute in a
least biased way a given number of bosonic secondaries, $\langle
n\rangle =\langle n^{(+)}\rangle + \langle n^{(-)}\rangle + \langle
n^{(0)}\rangle$, $\langle n^{(+)}\rangle =\langle n^{(-)}\rangle
=\langle n^{(0)}\rangle$. Using information theory approach (cf.,
\cite{IT}) their rapidity distribution was obtained in form of grand
partition function with temperature $T$ and chemical potential $\mu$.
In addition, the rapidity space was divided into {\it cells} of equal
size $\delta y$ each (it was fitted parameter). It turned out that
whereas the very fact of existence of such cells was enough to obtain
reasonably good multiparticle distributions, $P(n)$, (actually, in
the NB-like form), their size, $\delta y$, was crucial for obtaining
the characteristic form of the $2-$body BEC function
$C_2(Q=|p_i-p_j|)$ (peaked and greater than unity at $Q=0$ and then
decreasing in a characteristic way towards $C_2=1$ for large values
of $Q$, see Fig. \ref{fig:Fig2}) out of which one usually deduces the
spatio-temporal characteristics of the hadronization source
\cite{BEC} (see \cite{OMT} for more details). The outcome was
obvious: to get $C_2$ peaked and greater than unity at $Q=0$ and then
decreasing in a characteristic way towards $C_2=1$ for large values
of $Q$ one must have particles located in cells in phase space which
are of nonzero size. It means then that from $C_2$ one gets not the
size of the hadronizing source but only size of the emitting cell, in
\cite{OMT} $R\sim 1/\delta y$, cf. \cite{Z}. In the quantum field
theoretical formulation of BEC this directly corresponds to the
necessity of replacing delta functions in commutator relations by a
well defined peaked functions introducing in this way same
dimensional scale to be obtained from fits to data \cite{Kozlov}.
This fact was known even before but without any phenomenological
consequences \cite{Zal}.

Let us suppose now that we have mass $M$ and we know that it
hadronizes into $N=\langle n\rangle$ bosonic particles (assumed to be
pions of mass $m$) with equal numbers of $(+/-/0)$ charges and with
limited transverse momenta $p_T$. Let the multiplicity distribution
of these pions follows some NB-like form, broader than Poissonian
one. Suppose also that the two-particle correlation function of
identical particles, $C_2(Q)$, has the specific BEC form mentioned
above. How to model such process from the very beginning, i.e., in
such way that bosonic character of produced particles is accounted
for from the very beginning and not imposed at the end? We propose
the following steps (illustrated by comparison to some selected LEP
$e^+e^-$ data \cite{Data}):
\begin{itemize}
\item[$(1)$] Using some (assumed) function $f(E)$ select a particle
of energy $E^{(1)}_1$ and charge $Q^{(1)}$. The actual form of $f(E)$
should reflect somehow our {\it a priori} knowledge of the particular
collision process under consideration. In what follows we shall
assume that $f(E) = \exp\left( -E/T\right)$, with $T$ being parameter
(playing in our example the role of "temperature").

\item[$(2)$]Treat this particle as seed of the first {\it elementary
emitting cell} (EEC) and add to it, until the first failure, other
particles of the same charge $Q^{(1)}$ selected according to
distribution $P(E)=P_0\cdot f(E)$, where $P_0$ is another parameter
(actually it plays here the role of "chemical potential" $\mu =
T\cdot \ln P_0$). This assures that the number of particles in this
first EEC, $k_1$, will follow geometrical (or Bose-Einstein)
distribution, and in precisely this way one accounts for the bosonic
character of produced pions. This results in $C_2(Q)>1$ but only {\it
at one point}, namely for $Q=0$.

\item[$(3)$] To get the observed spread out of $C_2(Q)$ one has to
allow that particles in this EEC have (slightly) different energies
from energy of the particle being its seed. To do it allow that each
additional particle selected in point $(2)$ above have energy
$E^{(1)}_i$ selected from some distribution function peaked at
$E_1^{(1)}$, $G\left( E^{(1)}_1 - E^{(1)}_i\right)$.

\item[$(4)$] Repeat points $(1)$ to $(2)$ as long as there is enough
energy left. Correct in every event for every energy-momentum
nonconservation caused by selection procedure and assure that
$N^{(+)}=N^{(-)}$.
\end{itemize}
\begin{figure}[ht]
\hspace{1.55cm}
  \begin{minipage}[ht]{104mm}
    \centerline{
        \epsfig{file=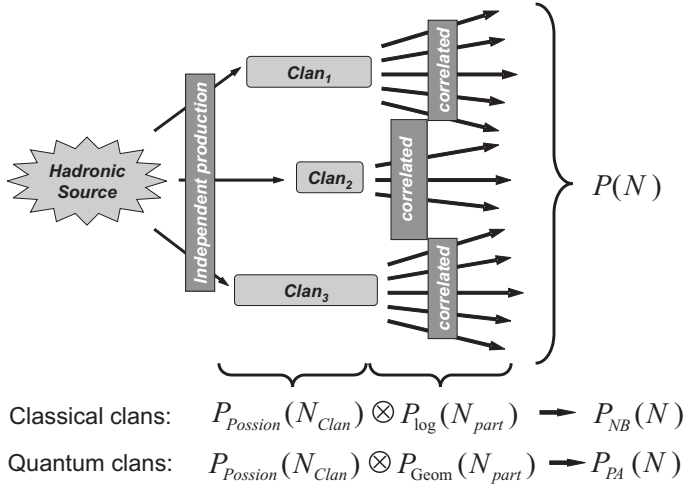, width=90mm}
               }
  \end{minipage}
  \caption{
\footnotesize {Schematic view of our algoritm, which leads to bunches
of particles ({\it clans}). Whereas in \cite{NBD} these clans could
consist of any particles distributed logarithmically in our case they
consist of particles of the same charge and (almost) the same energy
and are distributed geometrically to comply with their bosonic
character.}}
  \label{fig:Fig3}
\end{figure}
\begin{figure}[ht]
\begin{minipage}[ht]{57mm}
    \centerline{
        \epsfig{file=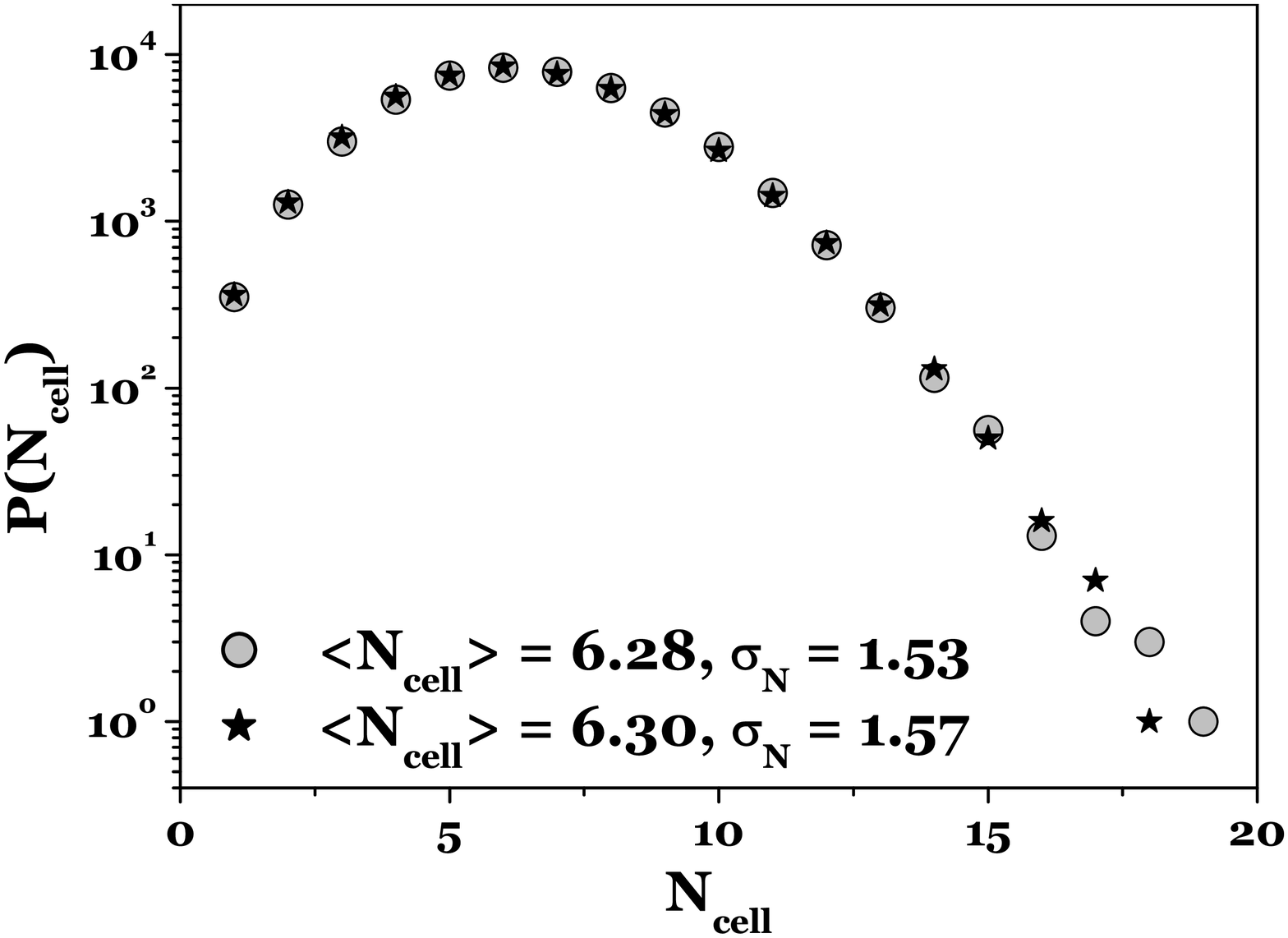, width=55mm}
     }
  \end{minipage}
\hfill
  \begin{minipage}[ht]{57mm}
    \centerline{
       \epsfig{file=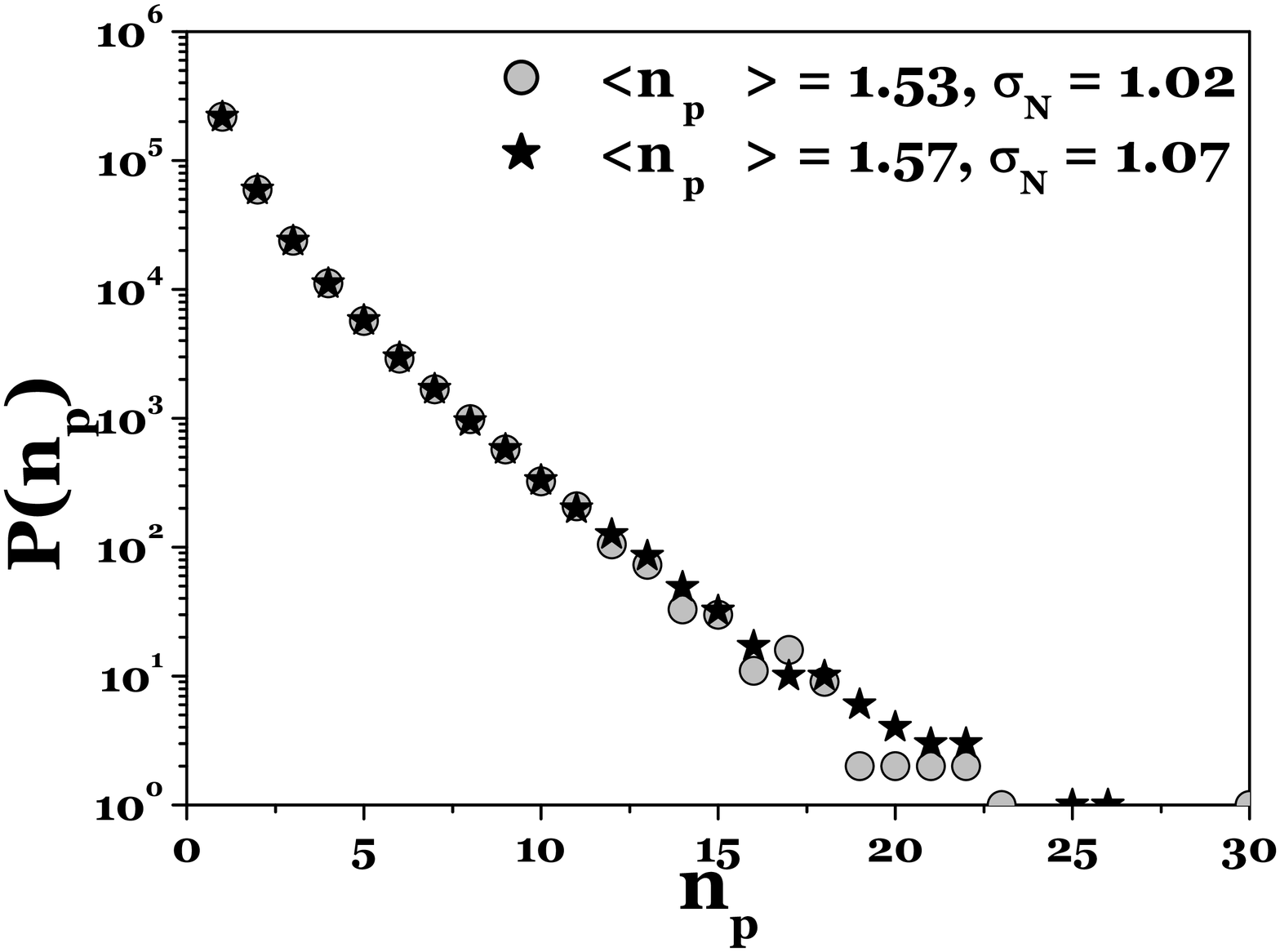, width=55mm}
     }
  \end{minipage}
  \hfill
  \begin{minipage}[ht]{57mm}
    \centerline{
        \epsfig{file=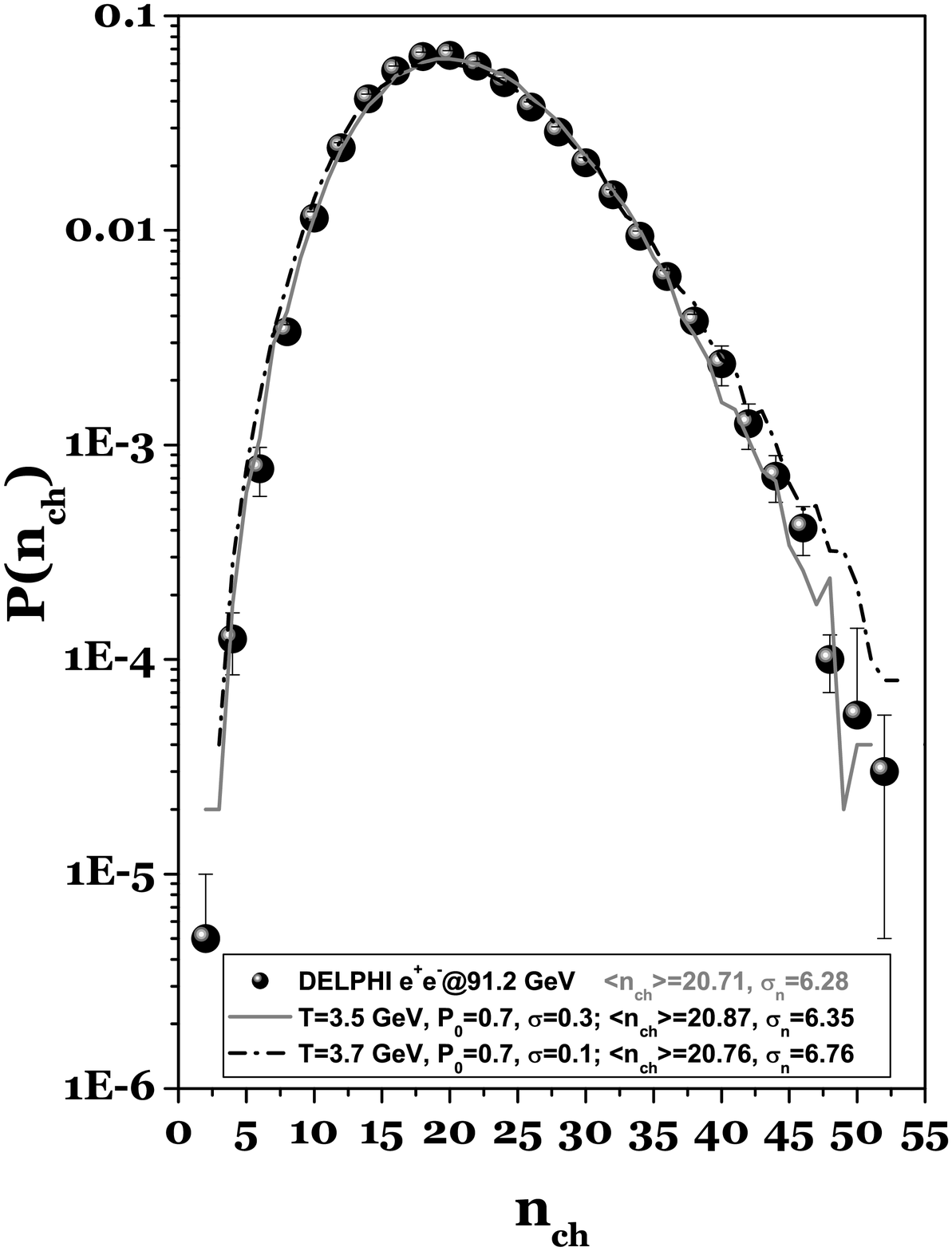, width=55mm}
     }
  \end{minipage}
\hfill
  \begin{minipage}[ht]{57mm}
    \centerline{
       \epsfig{file=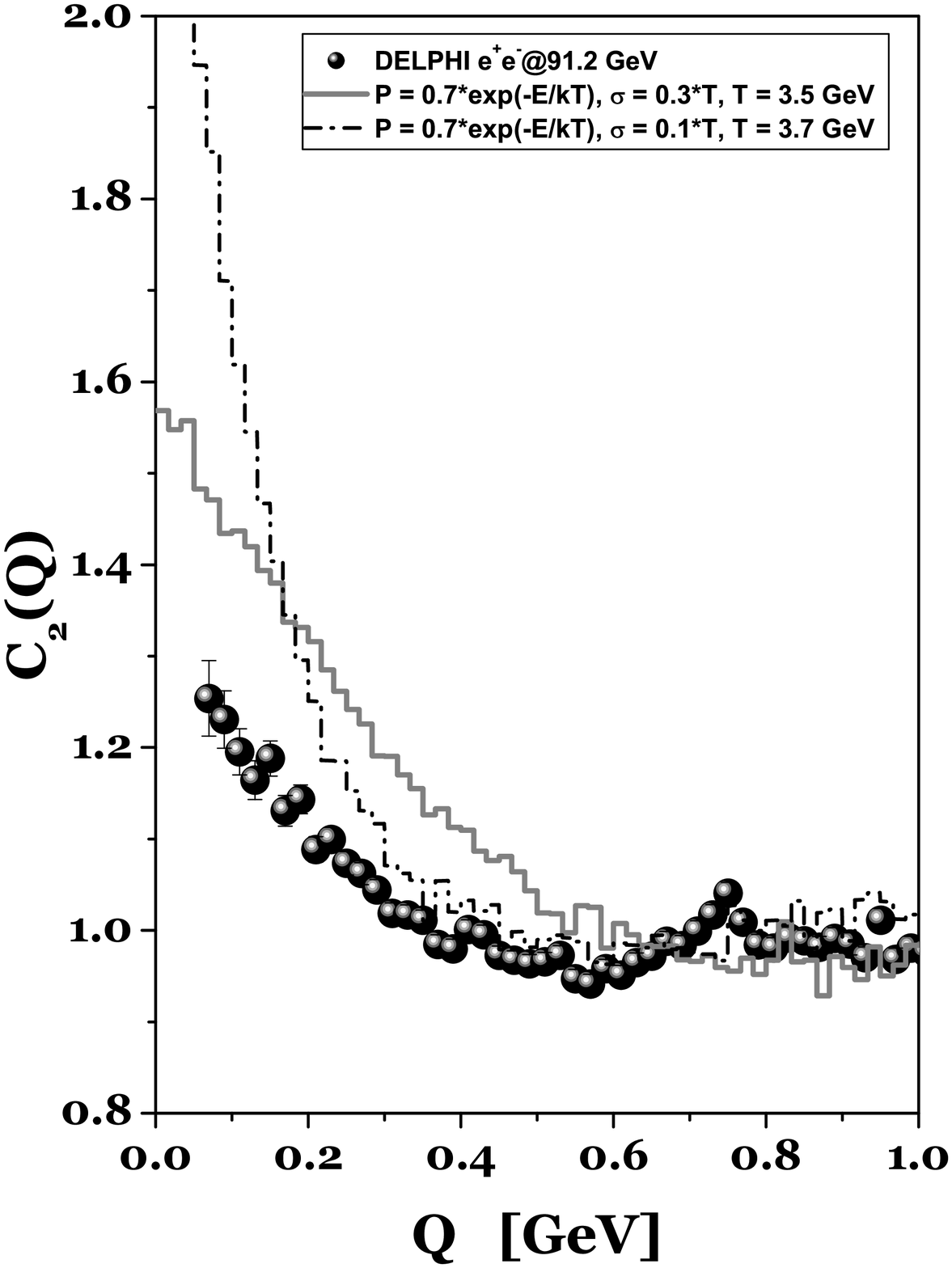, width=55mm}
     }
  \end{minipage}
  \caption{
\footnotesize {Upper panels: distribution of cells and particles in a
given cell. Lower-left panel: the corresponding summary $P(n)$ which
is convolution of both $P(n_{cell})$ and $P(n_p)$. Lower-right panel:
examples of the corresponding corresponding correlation functions
$C_2(Q)$ . Two sets of parameters were used. Data are from
\protect\cite{Data}.} }
  \label{fig:Fig4}
\end{figure}
As result we get a number of EECs with particles of the same charge
and (almost) the same energy, which we regard as being equivalent to
{\it clans} in the \cite{NBD} (see Fig. \ref{fig:Fig3}). These clans
are distributed in the same way as the particles forming the seeds
for those EEC, i.e., according to Poisson distribution (see Fig.
\ref{fig:Fig4}, upper-left panel). On the other hand, as was already
said, particles in each EEC will be distributed according to
geometrical distribution (see Fig. \ref{fig:Fig4}, upper-right
panel). As a result the overall distribution of particles will be of
the so called P\`olya-Aeppli distribution \cite{PA}. It fits our
examplatory data reasonable well. It is interesting to notice at this
point that to get NB distribution resulting from the classical clan
model of \cite{NBD} one should have logarithimc rather than
geometrical distribution of particles in EEC, which would then not
account for the bosonic character of produced secondaries. In this
respect our model differs from this {\it classical} clan model and we
see that what we have obtained is indeed its {\it quantum} version,
therefore its proposed name: {\it quantum clan model}.

The first preliminary results presented in Fig. \ref{fig:Fig4} are
quite encouraging (especially when one remembers that so far effects
of resonances and all kind of final state interactions to which $C_2$
is sensitive were neglected here). It remains now to be checked what
two-body BEC functions for other components of the momentum
differences and how they depend on the EEC parameters: $T$, $P_0$ and
$\sigma$. So far the main outcome is that BEC are due to EEC's only
and therefore provide us mainly with their characteristics (it is
worth to mention at this point that essentially this type of approach
has been also proposed to simulate Bose-Einstein condensate
phenomenon in \cite{Condens}).
\begin{figure}[ht]
\hspace{1.45cm}
  \begin{minipage}[ht]{104mm}
    \centerline{
        \epsfig{file=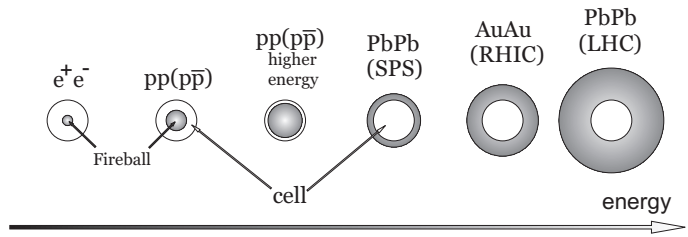, width=90mm}
               }
  \end{minipage}
  \caption{
\footnotesize {Schematic vie of how in our approach the size of the
EEC compares with the size of hadronizing fireball at different
energies and for different types of projectiles.}}
  \label{fig:Fig5}
\end{figure}
This should clear at least some of many apparently "strange" results
obtained from BEC recently (see Quark Matter 2004 proceedings,
especially \cite{QM2004}). The most intriguing is the fact that
apparently the "size" of the hadronizing source deduced from the BEC
data does not vary very much with energy and with the size of
colliding objects as has been naively expected \cite{BEC}. In our
approach this has simple explanation, see Fig. \ref{fig:Fig5}. The
point is that BEC are mainly sensitive to the correlation length,
which in our case is dimension of the emitting cell, not to dimension
of the "fireball" in which hadronization process takes place. The
size of this fireball depends mainly on the number of produced
secondaries \cite{QM2004}, which in our case is given by the
"partition temperature" parameter $T$ and by the "chemical potential"
parameter $\mu = T\cdot \ln P_0$. They are changing with mass $M$
(and therefore with the energy of reaction and the type of
projectile). On the other hand dimension of EEC is given entirely by
parameter $\sigma$ describing the spread of energy of particles
belonging to this EEC, which is only weakly depending on energy (if
at all). We shall close with mentioning that our approach accounts
also for multiparticle BEC and, because of this, by intermittence
effects seen in data (at least to some extend) \cite{UWW}.

\section*{Acknowledgements} Partial support of the Polish State
Committee for Scientific Research (KBN) (grant 2P03B04123 (ZW) and
grants 621/E-78/SPUB/CERN/P-03/DZ4/99 and 3P03B05724 (GW)) is
acknowledged.

\end{document}